\documentclass[runningheads]{llncs}
\usepackage[T1]{fontenc}

\usepackage{graphicx}
\usepackage{listings}
\usepackage{amsmath}
\usepackage{caption}
\usepackage{subcaption}
\usepackage{enumitem}
\usepackage{hyperref}

\begin{document}
\title{KetGPT -- Dataset Augmentation of Quantum Circuits using Transformers}
%
%

\author{Boran Apak\orcidID{0009-0000-3370-7473} \and
Medina Bandic\orcidID{0000-0003-4670-0988} \and
Aritra Sarkar\orcidID{0000-0002-3026-6892} \and
Sebastian Feld\orcidID{0000-0003-2782-1469}}
\authorrunning{B. Apak et al.}
%
\institute{
Quantum Machine Learning group, QuTech,\\ 
Department of Quantum \& Computer Engineering,\\
Delft University of Technology, The Netherlands
\email{boranapak1998@gmail.com,\{m.bandic,a.sarkar-3,s.feld\}@tudelft.nl}}
\maketitle              

\begin{abstract}
Quantum algorithms, represented as quantum circuits, can be used as benchmarks for assessing the performance of quantum systems. Existing datasets, widely utilized in the field, suffer from limitations in size and versatility, leading researchers to employ randomly generated circuits. Random circuits are, however, not representative benchmarks as they lack the inherent properties of real quantum algorithms for which the quantum systems are manufactured. This shortage of `useful' quantum benchmarks poses a challenge to advancing the development and comparison of quantum compilers and hardware. 

This research aims to enhance the existing quantum circuit datasets by generating what we refer to as `realistic-looking' circuits by employing the Transformer machine learning architecture. For this purpose, we introduce KetGPT, a tool that generates synthetic circuits in OpenQASM language, whose structure is based on quantum circuits derived from existing quantum algorithms and follows the typical patterns of human-written algorithm-based code (e.g., order of gates and qubits). Our three-fold verification process, involving manual inspection and Qiskit framework execution, transformer-based classification, and structural analysis, demonstrates the efficacy of KetGPT in producing large amounts of additional circuits that closely align with algorithm-based structures. Beyond benchmarking, we envision KetGPT contributing substantially to AI-driven quantum compilers and systems.


\end{abstract}

\section{Introduction}

The journey from knowledge and rule-based artificial intelligence to the contemporary era of data-driven deep neural networks-based machine learning~(ML) has marked significant milestones in artificial intelligence~(AI).
This type of AI, termed deep learning~(DL), focuses on recognizing and extracting patterns from vast datasets.
A proliferation of popular DL models and architectures contributed to use cases such as image and speech recognition, sequence prediction, and reinforcement learning.  
However, the application landscape changed dramatically with the emergence of generative models~\cite{harshvardhan2020comprehensive}, such as generative adversarial networks~(GAN) and variational autoencoders~(VAE). 
These models marked a profound shift in the capabilities of DL, allowing machines not only to recognize patterns in the data but also to generate new, coherent data that closely resembles the patterns learned from the training data.

Amid this diversity, the model that stands out in recent advances is the generative pre-trained transformer (GPT)~\cite{radford2018improving} based on the transformer architecture~\cite{vaswani2017attention}. 
Transformers achieve impressive performance on tasks like realistic text and code generation \cite{gpt4-paper,codegen} by capturing important information about the structure of sequences of data. 
GPT's ability to leverage massive scale with billions of parameters and self-supervised learning makes it the model of choice for natural language understanding and generation.
A wide spectrum of AI applications can be formulated as a language modeling and generation task, like chatbots, text summarization, question answering, code generation, medical diagnosis, and legal document review.

Simultaneously, another groundbreaking technology is being developed: quantum computers. 
Quantum computers can solve certain problems faster than classical computers~\cite{montanaro2016quantum} by employing information processing capabilities governed by the laws of quantum mechanics. 
To solve such problems, quantum algorithms, typically expressed as quantum circuits, need to be executed on quantum computers.
Besides serving the target use case, these circuits, defined in quantum assembly languages (QASM)~\cite{OPENQASM2.0}, are often also used to characterize, evaluate, and benchmark the quantum processors and related system software.
Moreover, system software, like the quantum compiler, often employs DL-based approaches to tackle the complexity of controlling large quantum processors.
This presents the need for large datasets of quantum circuits~\cite{qgym-paper,reinforcement-learning-paper} for the training of the ML-based quantum compilation passes, such as routing and mapping the circuits to a quantum processor.
However, at the moment, only a handful of quantum algorithms~\cite{QuantumZoo} are known to provide quantum computational benefits.
Due to the lack of large quantum circuit databases, these ML-based compilation techniques resort to randomly generated quantum circuits to train the model.
This use of unrepresentative training data can critically affect the performance of the quantum computer when deployed for pragmatic use cases. 

In an attempt to address this problem in quantum computing and inspired by the paradigm shift in language generation, in this work, we \textit{employ transformer models to generate realistic-looking quantum circuits to augment quantum circuit datasets.}

\vspace{0.7em}
\noindent This paper's contribution is threefold:

    \noindent 1. Introducing KetGPT, a transformer model capable of generating realistic-looking quantum circuits in the QASM language;
    
    \noindent 2. Developing a method to determine the quality of the generated QASM code using a different transformer model specifically designed for this task; and
    
    \noindent 3. Analyzing the generated circuits by extracting their structural parameters and comparing them to those of previously existing circuits.

\vspace{0.7em}
\noindent KetGPT can immediately be applied to the following use cases:

    \noindent $\bullet$ \textbf{Extending quantum circuit benchmarks datasets}: 
    KetGPT circuits offer a valuable expansion to existing circuit suites, such as those in \cite{quetschlich2023mqt,qbench-paper}, commonly employed for benchmarking and comparing quantum compilers and systems. Unlike typical synthetic circuits that consist of random gates on random qubits, KetGPT circuits emulate the behavior of real quantum algorithms, enhancing their relevance as benchmarks. Moreover, compared to the current practice of employing entirely random circuits with consistent width and depth, they present a compelling alternative for evaluating success metrics like quantum volume \cite{cross2019validating}. Given that quantum computers are designed to accelerate specific algorithms challenging for classical computers, assessing them using circuits that closely resemble these algorithmic structures is imperative. A dataset of KetGPT-generated quantum circuits is available as part of this software in Sec. \ref{code_and_dataset}.

    \noindent $\bullet$ \textbf{Automating quantum system software}: 
    Recent research uses machine learning models to enhance quantum compilation and error correction \cite{qgym-paper,reinforcement-learning-paper,overwater2022neural,machine-learning-mapper}. The substantial data required for training these models often leads researchers to resort to generating random circuits. However, a system that solves a certain problem should be trained on representative problem instances. Therefore, training a compiler to handle realistic circuits is more beneficial than training it on a random sample of gates, which makes KetGPT ideal for such a purpose \cite{qbench-paper}. In an ongoing project, KetGPT is being used to train a reinforcement learning agent for quantum circuit mapping on noisy quantum processors.

The remainder of this paper is structured as follows: The transformer models are introduced in Sec.~\ref{sec-Transformers}. Sec.~\ref{sec-KetGPT} introduces the main contribution of this work, KetGPT, a transformer model specifically designed to generate QASM files useful for benchmarking quantum system software. Additionally, a method is proposed to quantify how realistic these QASM files are. In Sec.~\ref{sec-results-and-discussion}, the generated code is examined and results are presented and discussed. Ultimately, Sec.~\ref{sec-conclusion-and-outlook} contains the conclusion of this work and presents suggestions for future work.

\section{Evolution and structure of Transformers}\label{sec-Transformers}
Transformer models, as introduced in the groundbreaking work  \cite{vaswani2017attention}, have changed the landscape of natural language processing. Their applications extend to code generation \cite{code-generation-using-gpt2,codegen} and music generation \cite{agostinelli2023musiclm}. Renowned for their proficiency in capturing dependencies within sequential data, these widely adopted machine-learning models have proven effective in various domains.

Before the advent of transformers, conventional models for natural language processing tasks, such as text generation, primarily relied on Convolutional Neural Networks (CNN) \cite{convolutional-NN}, Recurrent Neural Networks (RNN) \cite{RNN-paper}, and Long Short-Term Memory networks (LSTM) \cite{LSTM-paper}. However, these models encountered several challenges, including difficulties in handling long-range dependencies and a lack of parallelizability \cite{vaswani2017attention}. A transformer, on the other hand, is a highly parallelizable model, well-suited for training on extensive datasets, that excels at capturing longer-range dependencies and, therefore offers a significant improvement over earlier models.

In what follows, we review the three main components of the transformer model with quantum assembly language as the data.

\subsection{Tokenizer} \label{tokenisation-section}

\begin{figure}[t]
    \centering
\includegraphics[clip, trim=14.2cm 0cm 0cm 9.5cm,width=0.99\textwidth]{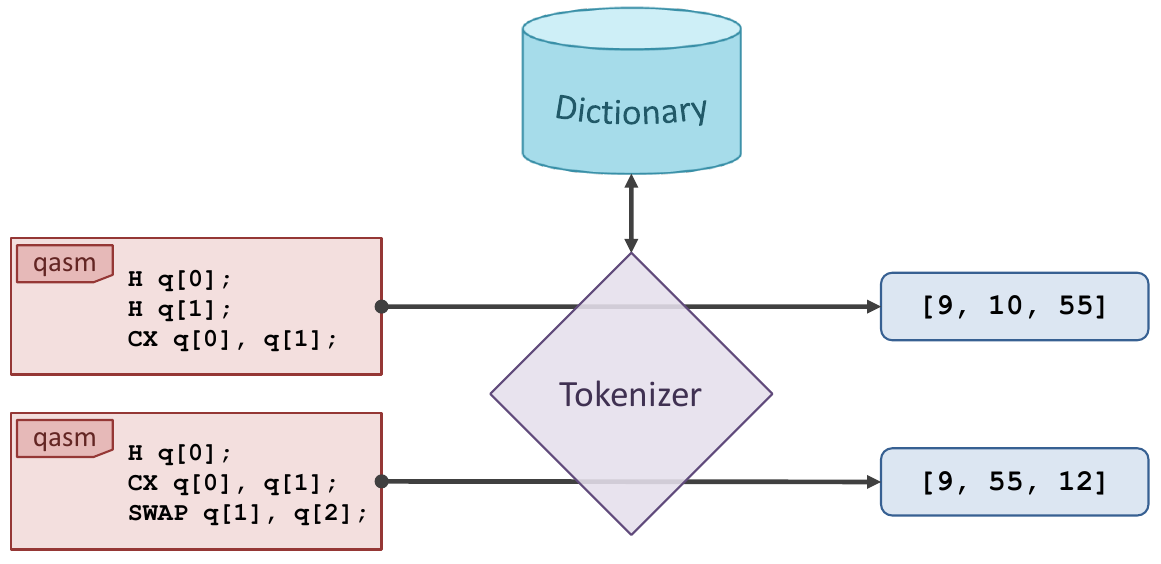}
    \caption{Tokenization Example. A sequence of QASM operations (in text file form) is provided as input, and each statement (a line of QASM code) is assigned to a number. The number assigned to each statement does not have an intuitive meaning; rather, it just depends on how the tokenization algorithm orders its vocabulary. Consequently, tokenizing a sequence of statements will create a list of numbers. It is important to note that both gate and qubit(s), we apply the gate on, matter for the assigned token. For instance, h q[0]; and h q[1]; would have different numbers assigned as shown.}
    \label{fig:tokenisation_example}
\end{figure}

It is well known that performing any kind of computations on strings necessitates converting them to numerical \textit{tokens} through a process called \textit{tokenization}. While this tokenization step is not explicitly outlined in the transformer architecture defined in \cite{vaswani2017attention} (as it falls under the domain of dataset preparation), it plays a crucial role in comprehending how information flows through a transformer model. A tokenizer plays a significant role in our case as using QASM code as input requires a different preprocessing type than with standard text. An example of the QASM code tokenization process is presented in Fig. \ref{fig:tokenisation_example}.

To fully describe a tokenization process, it is required to have a system for segmenting a sequence and a `dictionary' to establish the numerical association for each possible segment encountered using this segmentation system. There are different types of tokenization algorithms available.
For instance, instead of the scheme shown in Fig. \ref{fig:tokenisation_example}, every character can be converted to a number.
Thus, \verb|h q[0];| would be tokenized into 7 integers, one for each character and whitespace, instead of just a single token.

\subsection{Feed-forward neural network}\label{Feed Forward Neural Network section}

Neural networks \cite{goodfellow2016deep} play a key role in various machine-learning approaches and are one of the fundamental segments of transformer models. They consist of a series of layers that each perform a linear operation on the input followed by a (non-linear) activation function.  

To be precise, the value of each node in the network will be a linear combination of the values of the nodes in the previous layer weighted by the corresponding weights, passed through an activation function. Then a non-linear activation function (such as softmax \cite{bridle1989training} or ReLu \cite{GeLu-paper}) is applied so that the network can capture complex non-linear patterns.

A Feed-forward neural network is fully defined by specifying the number of layers, the number of nodes in each layer, the weights of every connection between nodes of a layer and a previous layer, a bias per node and the activation function per layer. To train a network, the desired architecture is initialized with (random) weights and biases. During training, the inputs are iteratively presented to the network and the weights and biases are adjusted to progressively align the network's output with the expected output for each specific input. This adjustment is typically done using a method called Stochastic Gradient Descent \cite{sgd-introduction-paper}. In this paper we are not focusing on the details of the neural networks, even though it represents the core of the transformer model, as it is widely and generally used as a base of most machine learning models. Instead, we will focus on the segments of the transformer that are specifically significant for our model, like \textit{self-attention}. 

\subsection{Self-attention} \label{Self-Attention-Section}

Self-attention is a mechanism that helps a transformer understand the relation between words and represents the main innovation in transformer models. Consider the sentence, ``The computer executes the program because it is told to.'' Humans effortlessly discern that ``it'' refers to the computer, not the program, but making automated systems distinguish this difference is very challenging. The inclusion of a self-attention component empowers transformers to establish such connections.

The input to the attention mechanism consists of queries, keys, and values. Each token in the input sequence corresponds to one query and key vector with dimension $d_k$ and a value vector with dimension $d_v$, but for computational purposes, the queries, keys and values for all tokens are packed into, respectively, matrices $Q$, $K$ and $V$. 
Thereafter, the main equation \cite{vaswani2017attention} describing the attention process is: 
\small
\begin{equation}
\label{eq:Attention_equation}
    \text{Attention}(Q, K, V) = \text{softmax}\left(\frac{Q K^T}{\sqrt{d_k}}\right) V,
\end{equation}
where softmax is the softmax function \cite{bridle1989training} and $K^T$ is the transpose of the $K$ matrix. 

The underlying idea of this equation is in the $QK^T$ term, representing the dot product between queries and keys to discern their ``inter-relation.'' Subsequently, this information forms an attention matrix akin to a correlation matrix. However, unlike a correlation matrix with values between $-1$ and $1$, the attention matrix adopts the form of a probability distribution, with values ranging from $0$ to $1$. The $\sqrt{d_k}$ scaling factor is there to obtain a more dimension-independent dot product, which helps train the network easier \cite{vaswani2017attention}. Multiplying this attention matrix with $V$ produces the final result, enriching the original matrix $V$ with insights into the inter-relations between queries and keys. For instance, elements with low scores in the attention matrix, close to $0$, are drowned out. To illustrate, in the context of encoding the sentence ``The computer executes the program because it is told to.''  represented by matrices $Q$, $K$, and $V$, the operation $\text{Attention}(Q,K,V)$ returns a matrix that embodies this sentence with information about the inter-relations between the words (e.g., clarifying that ``it'' refers to the computer and not the program).

\section{KetGPT - Transformers for quantum circuit generation}\label{sec-KetGPT}

\begin{figure}[htb]
    \centering
    \includegraphics[clip, trim=3.5cm 0.5cm 3.5cm 0.5cm,width = 0.92\textwidth]{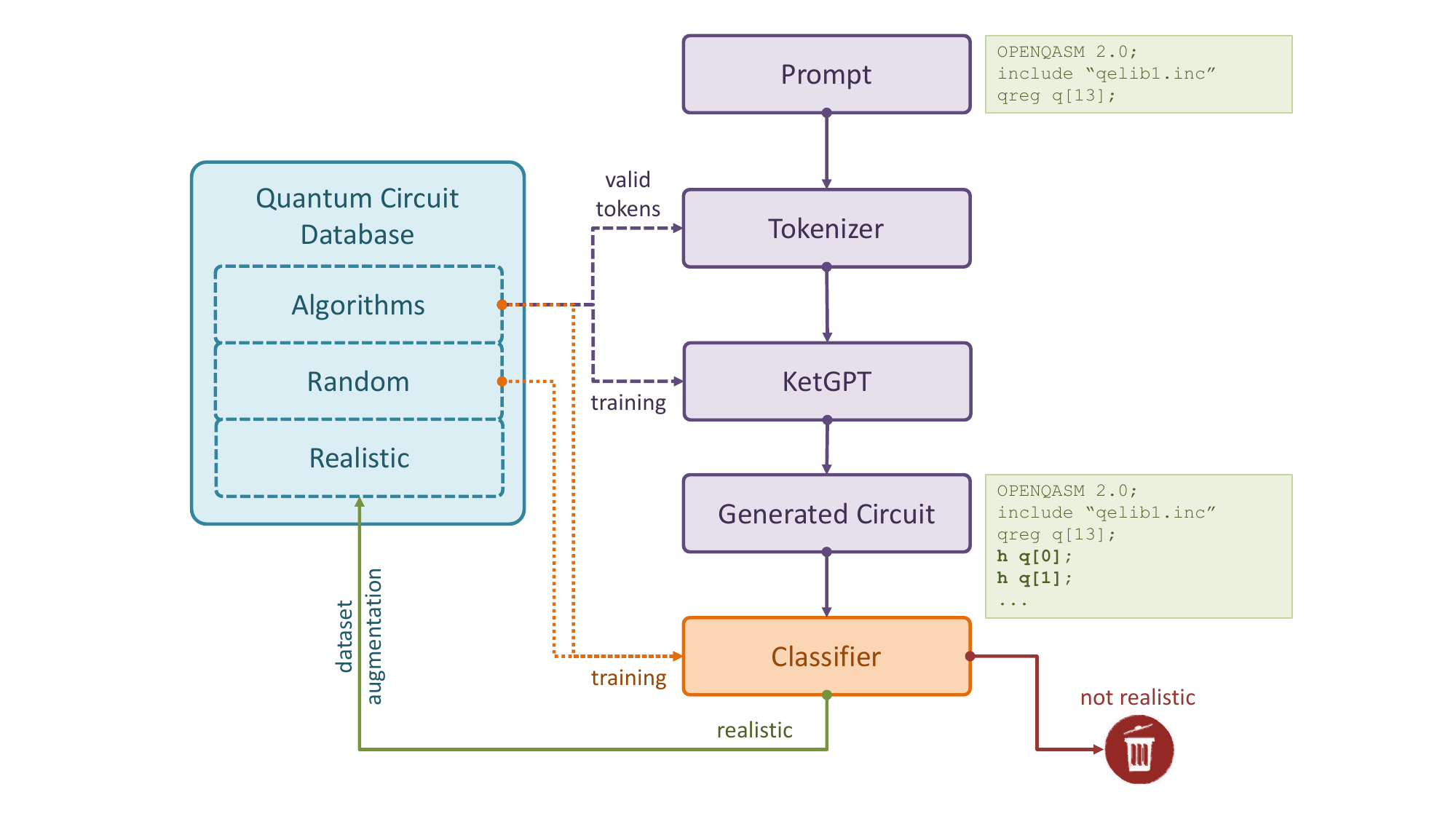}
    \caption{KetGPT Workflow: Firstly, a given text prompt is tokenized. These tokens are fed into the KetGPT model, which was trained with quantum circuits from an existing quantum circuit database. KetGPT then generates text to continue the given prompt, yielding a synthetic circuit. A separate transformer classifier model, trained to distinguish real from random quantum circuits, tests if the generated circuit is realistic. If the test is positive, it can be used to augment the quantum circuit database.}
    \label{fig:blocks}
\end{figure}

This section presents KetGPT, a novel software tool designed to generate quantum algorithm-based circuits. These circuits can serve as essential benchmarks for evaluating the performance of both existing and forthcoming quantum systems. Within this section, we delve into the technical intricacies of KetGPT, offering a comprehensive understanding of its architecture and methodology. Fig. \ref{fig:blocks} shows an overview of the KetGPT design and overall workflow. 

\subsection{Input dataset and data preprocessing} \label{qubit-count-list}

Several datasets of quantum circuits suitable for benchmarking are available \cite{QASMbench,Revlib,qbench-paper}, including MQT Bench \cite{quetschlich2023mqt}, which is utilized in this study. QASM files were generated to depict circuits implementing algorithms spanning 2 to 100 qubits, employing OpenQASM 2.0 \cite{OPENQASM2.0}. In cases where algorithms were incompatible with a specific qubit count, such as those requiring an uneven number of qubits, all valid circuits within the feasible range were generated. The full dataset and additional details can be found in Sec. \ref{code_and_dataset}.

The files taken from the dataset require preprocessing in order to comply with the transformer model. This involves making minor adjustments to the QASM files in the dataset (e.g., removing comments). Due to technical constraints -- specifically, the model's incapacity to process large files -- a maximum circuit length of 1024 QASM statements is enforced. 
This limitation is specific to the hardware's RAM constraints and not a general technical restriction. Following the preprocessing step, the final dataset comprises 713 QASM files.

\subsection{Generator: architecture and tokenizer} \label{section_tokenisation_generator}

When it comes to generating text and code, a decoder-only transformer architecture \cite{code-generation-using-gpt2} is a popular choice. Accordingly, for the generation of QASM files, we have opted for the GPT-2 model architecture \cite{GPT-2-paper}, known for its use of a decoder-only transformer. The Python code to construct this architecture is openly accessible through the GPT-2 implementation in the Hugging Face ``Transformer'' python library \cite{Huggingface-Transformers,huggingface2021transformers}.

As discussed in Sec. \ref{tokenisation-section},  we employ a \textit{tokenization} approach to transform the dataset text into tokens. The original implementation of GPT-2 relies on a form of tokenization known as Byte Pair Encoding (BPE). To comprehend this method intuitively, it dissects text into components (e.g., `training' into `train' and `ing'), facilitating a better grasp of the full word's meaning. However, a drawback is that it may allow the generation of QASM code that is not syntactically correct, such as the potential generation of the line ``hh q0q1;''. To address this, we modified the tokenization method for the generator to only permit syntactically correct QASM code as tokens. This modification was implemented by adjusting the GPT2Tokenizer class. By compiling a list of all valid QASM statements in the dataset and using it as our vocabulary, we ensure that any token generated by the model will be a valid QASM statement. The generator model workflow consists of the following four parts:

\textbf{Preparation:} The process of generating tokens using the generator model unfolds as follows: i) A list is compiled containing the qubit count for every circuit in the dataset, along with another list containing the number of gates for each circuit; ii) From these lists, a qubit count and a number of gates are randomly selected, establishing the parameters for the QASM file to be generated; and finally, iii) With these parameters in hand, any invalid QASM statement related to the selected qubit count is filtered out. For instance, if the chosen qubit count is 5, all gates involving qubit 13 are disregarded. This is achieved by preventing the generator model from producing these tokens.

\textbf{Model input:}
The model will receive input as the following: 
\small
\renewcommand{\baselinestretch}{0.8}
\begin{verbatim}
        OPENQASM 2.0;
        include "qelib1.inc"
        qreg q[{}];
\end{verbatim}

where \{\} will contain the chosen qubit count. This is the way all the QASM files in our dataset start, and it gives us an opportunity to control the qubit count in a simple manner. 

\textbf{Generation scheme:} Whenever a new probability distribution over the tokens is generated, the top-k strategy \cite{top-k-paper} is employed, where the $k=5$ most probable tokens are identified. From this subset, a new token is selected based on the renormalized probability distribution over these five tokens (the renormalization ensures that all probabilities add up to one). This approach introduces additional randomness into the QASM file generation process while maintaining the realism of the generated tokens, as the five most probable tokens are typically viable candidates. Furthermore, it is specified that sequences of 15 tokens should not repeat within the file. While this constraint may not align perfectly with QASM code generation, in which algorithms often contain repetitive sequences, it serves to prevent instances where the transformer model becomes stuck in a loop, repeatedly predicting the same sequence.
The top-k generation process iterates until the desired number of gates is reached.

\textbf{Post-processing:}
Finally, to guarantee the validity of all generated files, all quantum and classical registers utilized in the generated file are instantiated at the beginning of the QASM file. This ensures every file, including its header, is syntactically correct.

\subsection{Verification method: KetGPT classifier}

Once the generator produces the QASM files, the next step is to assess their authenticity. To determine whether the generated QASM files exhibit a ``realistic'' quality, we employed a binary classifier. This classifier's task is to distinguish whether a generated QASM file bears a closer resemblance to files from our algorithm-based circuit dataset or aligns more with a randomly generated QASM file \cite{qbench-paper}.

The classifier adopts an encoder-only transformer model, specifically the architecture of the DistilBERT model \cite{Sanh2019DistilBERTAD}, leveraging the implementation from the Huggingface transformers library \cite{huggingface2021transformers}. This model is a smaller version of the highly influential encoder-only BERT model \cite{BERT-paper} and is chosen for quicker training and inference.

Unlike the generator, which required a customized tokenization method to ensure the generation of valid QASM code, the classifier employs the \textit{tokenization} method used to train the original DistilBERT model, known as WordPiece \cite{wordpiecepaper}. This method, similar to the BPE tokenizer briefly mentioned in Sec. \ref{section_tokenisation_generator}, breaks down words into sub-words. It is important to note that the choice of how these sub-words are determined distinguishes WordPiece from BPE, but this is not pertinent to this work. To adapt the QASM sequences for the classifier, the tokenization truncates them after 512 tokens. Since these tokens represent sub-words instead of complete QASM lines, the 512-token limit corresponds to approximately 50 lines of QASM code, dependent on the sequence. This adjustment ensures compatibility with the maximum input size of the classifier model used. While this approach has the drawback of only considering the initial portion of the QASM file in determining its authenticity, it offers the advantage of expedited training and inference, necessitating a less technically intricate model. Moreover, the initial segment of a QASM file typically provides sufficient cues to discern its nature as random or structured.

During the \textit{training} phase of the classifier, a dataset is prepared in which all real quantum circuits are assigned the label `0' (total of 1112 QASM files). Correspondingly, an equal number of QASM files are randomly generated, comprising gates randomly selected from a list of all unique QASM statements in the dataset, and labeled `1'. To ensure fairness in the classification process, akin to the methodology employed for generating KetGPT QASM files, the randomly generated QASM files are structured to encompass the same distribution of qubit counts and number of gates as the original dataset. 
Subsequently, the model is trained on the labeled dataset, and upon completion of training, the trained model is employed to predict whether the KetGPT-generated circuits are classified as `0' or `1', indicating their proximity to genuine algorithms or random circuits, respectively.

\subsection{Implementation details} \label{section-hardware-implementation}

Our experiments were conducted using a Jupyter notebook \cite{jupyternotebook} executed on the \textit{Google Colab} environment \cite{GoogleColab}. This Notebook is provided in Section \ref{code_and_dataset}. The Google Colab GPU has 16Gb of GDDR6 memory, 320 Turing tensor cores and 2560 CUDA cores. At the time of writing, Google Colab uses Python version 3.10.12. Relevant packages for the code used to obtain the results of this work are the transformers \cite{Huggingface-Transformers} (version 4.34.0) and datasets \cite{Huggingface-datasets} (version 2.14.5) libraries from Huggingface, PyTorch \cite{pytorchcitation} (version 2.0.1+cu118) and NumPy \cite{NumPy} (version 1.23.5).

Tab. \ref{tab:generator_model_settings} contains the parameters that define the structure of our generator model. Default values correspond to those used in the original GPT-2 implementation \cite{GPT-2-paper}. The training settings are specified in Tab. \ref{tab:training_settings}. On the other hand, Tab. \ref{tab:classifier_model_settings} specifies the settings that were used to define the \textit{classifier model}.  
The training settings for the classifier model are in Tab. \ref{tab:training_settings_classifier}. All the parameters' detailed definitions can be found in \cite{huggingfaceGPT2}.

\begin{figure}
    \centering
    \begin{minipage}{0.45\textwidth}
        \centering
        \captionof{table}{Generator model settings}
        \begin{tabular}{|c|c|}
            \hline
            \textbf{Name} & \textbf{Value} \\
            \hline
            n\_embd & 768 (default) \\
            \hline
            n\_layer & 3 \\
            \hline
            n\_head & 4 \\
            \hline
            n\_positions & 1024 (default) \\
            \hline
            vocab\_size & 48291 \\
            \hline
        \end{tabular}
        \label{tab:generator_model_settings}
    \end{minipage}
    \hfill
    \vspace{0.3cm}
    \begin{minipage}{0.45\textwidth}
        \centering
        \captionof{table}{Generator training settings}
        \begin{tabular}{|c|c|}
            \hline
            \textbf{Name} & \textbf{Value} \\
            \hline
            Epochs & 5 \\
            \hline
            Learning Rate & 5e-5 (default)\\
            \hline
            Batch Size & 4 \\
            \hline
            Optimiser & AdamW (default)\\
            \hline
            Loss function & Cross-entropy (default)\\
            \hline
        \end{tabular}
        \label{tab:training_settings}
    \end{minipage}
    \hfill    
    \begin{minipage}{0.45\textwidth}
        \centering
        \captionof{table}{Classifier model settings}
        \begin{tabular}{|c|c|}
            \hline
            \textbf{Name} & \textbf{Value} \\
            \hline
            n\_embd & 768 (default) \\
            \hline
            n\_layer & 6 (default) \\
            \hline
            n\_head & 12 (default)\\
            \hline
            n\_positions & 512 (default) \\
            \hline
            vocab\_size & 30522 \\
            \hline
            
        \end{tabular}
        \label{tab:classifier_model_settings}
    \end{minipage}
    \hfill
    \vspace{0.3cm}
    \begin{minipage}{0.45\textwidth}
        \centering
        \captionof{table}{Classifier training settings}
        \begin{tabular}{|c|c|}
            \hline
            \textbf{Name} & \textbf{Value} \\
            \hline
            Epochs & 3 \\
            \hline
            Learning Rate & 5e-5 (default)\\
            \hline
            Batch Size & 4 \\
            \hline
            Optimiser & AdamW (default)\\
            \hline
            Loss function & Cross-entropy (default)\\
            \hline
        \end{tabular}
        \label{tab:training_settings_classifier}
    \end{minipage}
\end{figure}

 It is worth noting that KetGPT training time was 240 seconds, and generating 1000 QASM files took 8818 seconds (147 minutes), or 8.8 seconds per generated file on average. However, the QASM files are of varying size (as explained in Sec. \ref{qubit-count-list}), and the amount of time needed to generate one file is dependent on its size, so this number should be taken as a rough estimate.

\section{Results and discussion}\label{sec-results-and-discussion}

In this section, we unveil outcomes of this work by showing the results of the three verification steps: manual inspection and Qiskit execution, transformer-based classifier and structural analysis of the circuits. Note that the usage of the term 'realistic'  or 'real' when describing the circuits generated by KetGPT is not meant to be interpreted as describing circuits that implement useful quantum algorithms. The circuits might describe some undiscovered quantum algorithms, but it is nearly impossible to reverse engineer an explainable description.

\subsection{Manual inspection}

\begin{figure*}[h]
    \centering
    \begin{subfigure}{0.23\textwidth} 
        \centering
        \includegraphics[width=\textwidth]{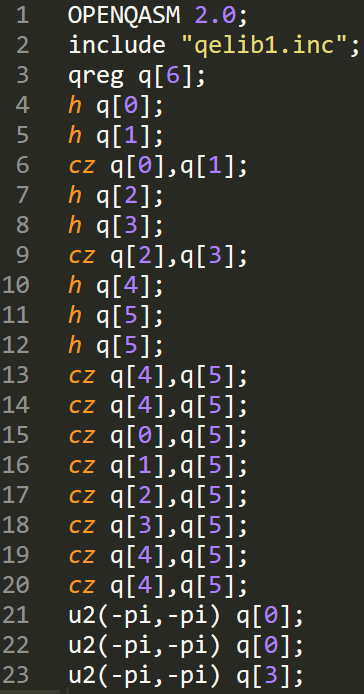}
        \caption{KetGPT}
        \label{fig:qasm-code-KetGPT}
    \end{subfigure}
    \hfill 
    \begin{subfigure}{0.26\textwidth} 
        \centering
        \includegraphics[width=\textwidth]{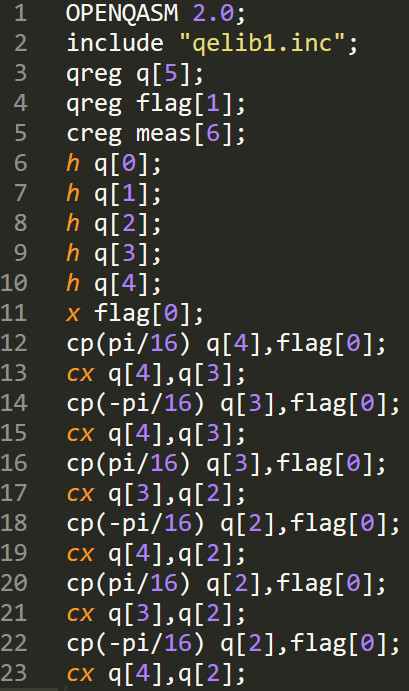}
        \caption{Real}
        \label{fig:qasm-code-Real}
    \end{subfigure}
        \hfill 
    \begin{subfigure}{0.455\textwidth} 
        \centering
        \includegraphics[width=\textwidth]{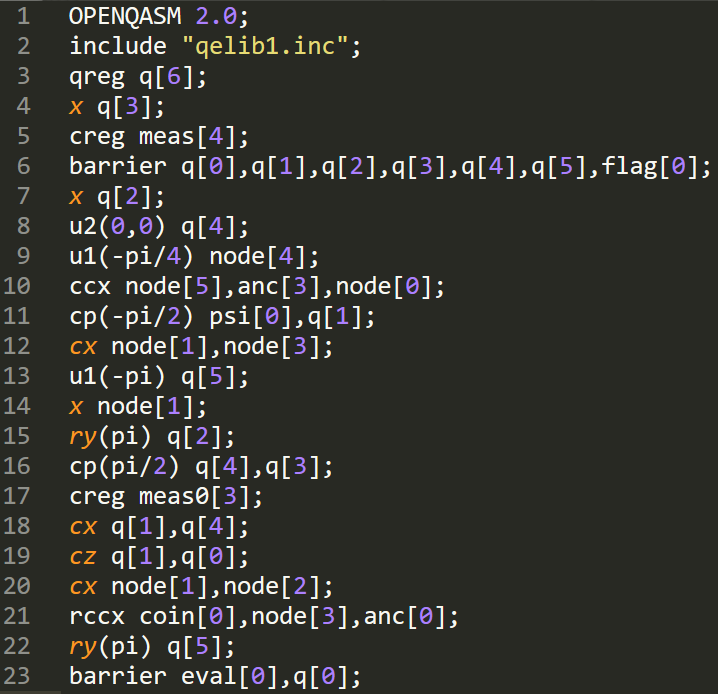}
        \caption{Random}
        \label{fig:qasm-code-Random}
    \end{subfigure}
    \caption{Side-by-side comparison between the lines of a 6 qubit QASM file generated by KetGPT (a), algorithm-based circuit (b) and a random circuit(c).} 
    \label{fig:Comparison-KetGPT-Real-Random} 
\end{figure*}

First, we manually examine the QASM lines of a circuit produced by KetGPT. We juxtapose this with the initial lines of both a genuine and a completely random circuit to establish a comparative analysis. One can observe some patterns shown in the files of Fig. \ref{fig:Comparison-KetGPT-Real-Random}: The lines within the KetGPT file and the real file exhibit structured patterns, such as the repetition of Hadamard and 2-qubit gates (CX and CZ), whereas the fully random circuit lacks such repetitive sequences. Additionally, it is noteworthy that the order in which the Hadamard gates are applied in the KetGPT and the real circuit follows an ascending order based on qubit numbers, whereas in the fully random circuit, as expected, there is no logical order of operations. Importantly, the random circuit includes invalid statements, such as operations on nodes that were never defined (e.g., an operation on node 4 is instructed, but node 4 was never defined). However, this error is also occasionally present in files generated by KetGPT, albeit seemingly less frequently. The fact that it is not specifically forbidden for KetGPT to generate invalid statements, but it still generates such statements considerably less often than random files, can also be seen as a realistic feature of KetGPT-generated data. Note that we also ran all our circuits within the Qiskit framework \cite{Qiskit} where 96\% of the circuits passed the compilation process successfully.

Based on the provided examples and the illustration in Fig. \ref{fig:Comparison-KetGPT-Real-Random}, a visual examination strongly indicates that KetGPT-generated circuits exhibit characteristics reminiscent of real quantum circuits. This observation underscores the promise of employing transformers to generate quantum circuit data.

\subsection{Classifier-based evaluation}\label{section-data-leakage}

As a second measure of verification, we developed and trained a classifier model to determine whether KetGPT circuits resemble more real algorithm-based or random quantum circuits. As input, we created a dataset with the same amount of real and random circuits (1112 each) and used 85\% of the data for training and 15\% for testing the classifier.
\begin{figure}
    \centering
        \includegraphics[width=0.5\textwidth]{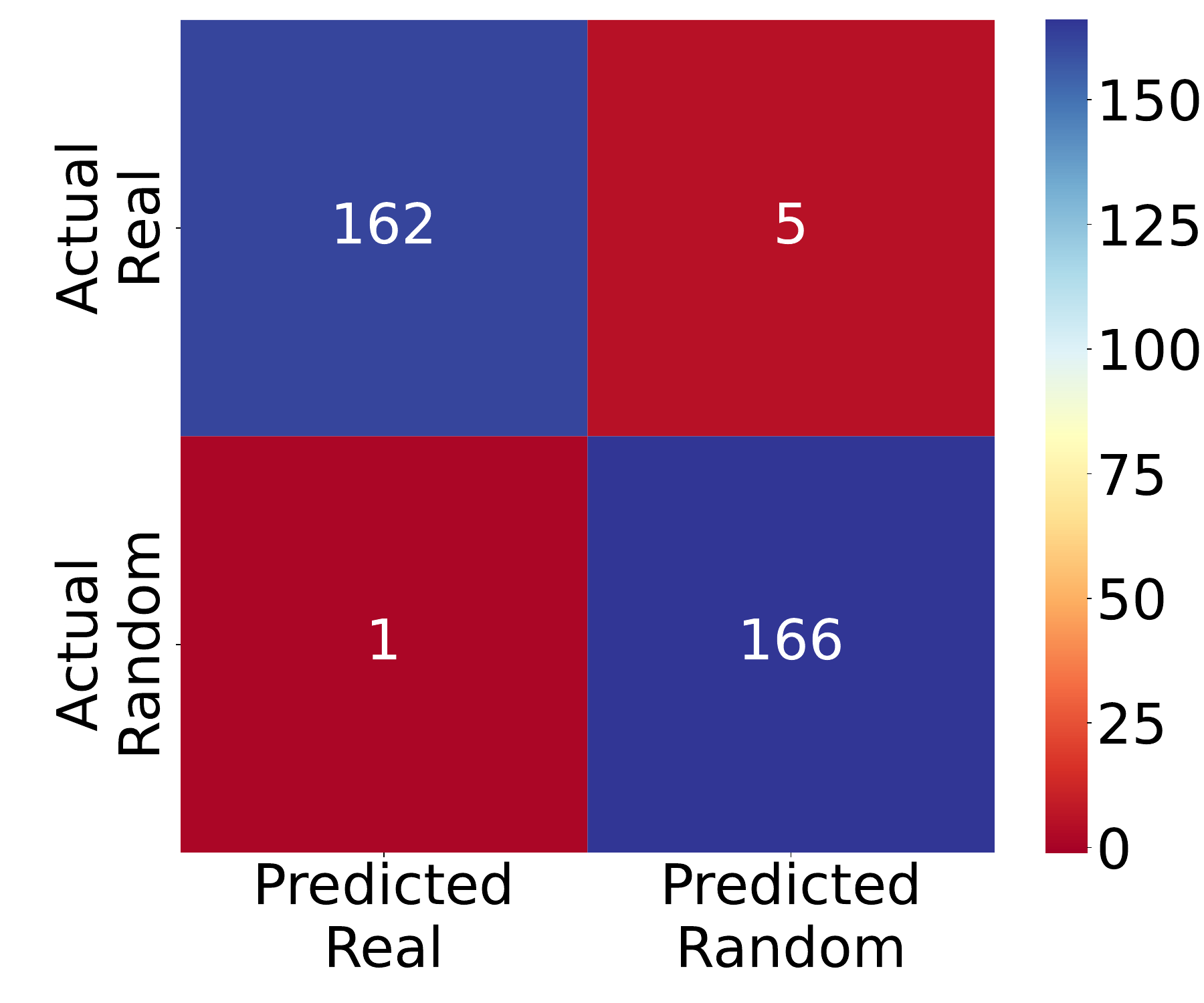}
        \caption{Classifier performance on a test dataset illustrated by a confusion matrix. Diagonal values of the matrix are correctly predicted: only 5 QASM files that are actually ``Real'' are predicted as being ``Random'', and 1 QASM file that is `Random' is predicted as being `Real'.}
        \label{fig:confusion matrix 6 epochs}
\end{figure}

To assess the model's performance, a confusion matrix is employed to ascertain the alignment between the model's predictions and the actual labels of the data. The corresponding confusion matrix for this evaluation is depicted in Fig. \ref{fig:confusion matrix 6 epochs}. A total of 328 out of 334 test dataset values are predicted correctly, which means that the classifier model achieved an accuracy of 98.2\%.

Subsequently, the classifier was tasked to classify 1000 KetGPT QASM files as either more similar to its training dataset (real algorithm-based) or to completely random qauantum circuits. Among the 1000 circuits evaluated, 999 were classified as authentic, indicating a classification accuracy of 99.9\%.

It is difficult to evaluate the reliability of the model. The high accuracy could potentially be explained by the fact that the test dataset consists of a random subset of the total data. It is possible, for instance, that the Deutsch-Jozsa algorithm on 6 qubits is part of the training dataset, and Deutsch-Jozsa on 5 qubits is in the test dataset. The similarity between the training and testing data may influence the accuracy metric calculation. Nonetheless, using different instances of the same algorithms for the datasets was inevitable due to the limited availability of diverse algorithms. The random QASM files in the test set, however, are not similar to the random files in the training dataset, and are still predicted correctly every time.

Taking all of these considerations into account, including the classifier's accuracy when evaluated, it appears that the classifier is capable of discerning realistic features within the data. However, determining whether this proficiency results from the model overfitting to specific features of QASM files or genuinely learning relevant aspects of realistic circuits presents a challenge.

\subsection{Analysis based on circuit structure}

\begin{figure}[bht]
    \centering
    \includegraphics[clip,trim=8.5cm 4cm 2.5cm 4cm ,width = 0.7\textwidth]{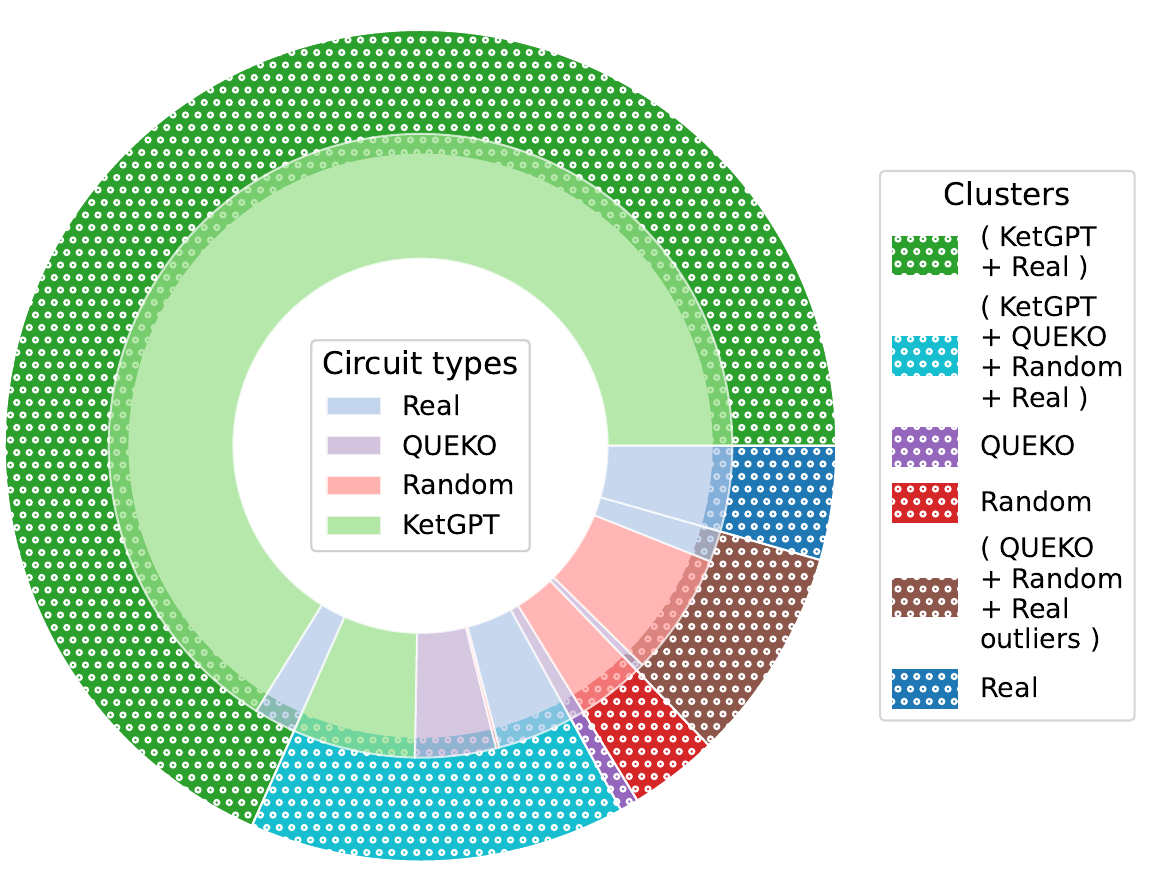}
    \caption{
    The distribution of clusters obtained through structural parameters analysis is depicted. Each segment in the outer circle represents clusters characterized by the same types of circuits (e.g., the dark green segment encompasses all clusters that consist of KetGPT and real circuits). The inner circles display the quantity of each circuit type within the respective outer circle segment.}
    \label{fig:clustering_result}
\end{figure}

Another approach to quantifying and validating KetGPT involves extracting structural parameters from circuits. Within this approach, a circuit is transformed into interaction and gate dependency graphs \cite{qbench-paper} and then analyzed based on quantum compilation-related, graph theory-based (e.g., degree of nodes) parameters. Following this methodology, we extracted the suggested 23 metrics \cite{bandic2024} from our KetGPT circuits in order to compare them with existing circuit dataset. For comparison, we followed another method suggested in \cite{qbench-paper} and clustered the circuits (KetGPT and qbench \cite{qbench-paper} circuits) based on the extracted parameters to discover groups of ultimately structurally similar circuits. The benchmark set \textit{qbench} consists of real algorithm-based circuits, random circuits, and QUEKO circuits (synthetic circuits with predefined depth and gate count)\cite{Queko}, so by doing clustering, we could see where KetGPT would belong within these groups, or if it would form its own.
Notably, we refrained from utilizing this benchmark set for creating KetGPT circuits, ensuring an unbiased evaluation.

Clustering is done in a two-level manner: first based on size and then sub-clusters based on the structure of the quantum circuits, resulting in a final tally of 18 clusters. For clarity, we consolidated clusters sharing identical circuit structures (in terms of circuit types) into one and illustrated the distribution in Fig. \ref{fig:clustering_result}. The depicted clustering reveals that KetGPT circuits consistently align with real circuits and never with completely random ones. Additionally, a smaller portion of QUEKO circuits exhibit a similar association with both KetGPT and real circuits. Given that QUEKO circuits aim to mimic realistic behaviors more closely than classical random circuits \cite{Queko}, this observation is logical. Fig. \ref{fig:clustering_result} also suggests how much KetGPT contributes to having more realistic circuits in the whole set (green segments of the inner circle).

\section{Conclusion and outlook}\label{sec-conclusion-and-outlook}
The scarcity of quantum circuits 'useful' for benchmarking, stemming from limitations in existing datasets, poses a significant challenge to the progress of quantum compiler and hardware development. To address this gap, our research introduces KetGPT, a tool that utilizes the Transformer machine learning architecture to generate synthetic circuits resembling real-world quantum algorithms. We verified our resulting circuits three-fold by: 1) Running the circuits with Qiskit framework and manual inspection, we achieved a 96\% success rate (without error or warning); 2) Implementing and training a transformer-based classifier for distinguishing between 'real' and random algorithms which classified KetGPT circuits as real in 99\% of the cases; and 3) Characterizing the generated circuits by extracting structure-based properties and clustering them together with another dataset containing real and random circuits. The analysis revealed that all our circuits closely resembled the structure of algorithm-based ones, and showcased the expansion of the dataset. In conclusion, this three-step, extensive verification shows that KetGPT can augment realistic and executable quantum circuit dataset(s).

Our future steps in expanding and improving KetGPT include: i) Exploring alternative generation schemes, such as top-p \cite{top-p-paper}, beam search \cite{beam-search-paper}, or contrastive search \cite{contrastive-search-paper}, to compare their effectiveness in generating QASM files or, development of a generation scheme tailored specifically for QASM file generation; ii) Reconsidering the representation of QASM statements as discrete tokens: Introducing an arbitrary gate token to accommodate QASM files with arbitrary angles, using a transformer trained for this purpose in post-processing; and iii) Modifying the tokenization scheme by separating gates and target qubits into distinct tokens (e.g., treating `Hadamard gate' and `on qubit 1' as separate tokens) and ensuring that the adjusted scheme generates only valid QASM expressions and exploring its scalability for higher qubit counts.

In summary, we are confident that KetGPT holds the promise to not only significantly influence the benchmarking of quantum systems, but also to serve as a valuable input for data-intensive, AI-based solutions in the development of innovative quantum compilers and systems.

\section{Software Availability} \label{code_and_dataset}

The code that was used for this work is provided as a Jupyter notebook \cite{jupyternotebook}, which was executed in the Google Colab environment \cite{GoogleColab}, available at:

\noindent\href{https://colab.research.google.com/drive/1dbtJX6q8sED4yrb1I09KUuXWYH0AVN8r}{https://colab.research.google.com/drive/1dbtJX6q8sED4yrb1I09KUuXWYH0AVN8r}.

The data that was used for this work, comprising of the training dataset, and a KetGPT folder that contains: the pre-trained KetGPT model, the KetGPT tokenizer, the pre-trained classifier model, all KetGPT generated circuits and all random circuits, is available at:
\noindent\href{https://www.kaggle.com/datasets/boranapak/ketgpt-data}{https://www.kaggle.com/datasets/boranapak/ketgpt-data}.

\section{Acknowledgments}\label{sec-acknowledgements}
MB and SF acknowledge funding from Intel Corporation. AS acknowledges funding from the Dutch Research Council (NWO).

\bibliographystyle{splncs04}
\bibliography{ref}
%

\end{document}